\documentclass[12pt,letterpaper]{article}

\usepackage{amsmath}
\usepackage{amsfonts}
\usepackage{amssymb}
\usepackage{latexsym}
\usepackage{graphicx}
\usepackage{float} 
\usepackage{bm}

\usepackage{dsfont}
\usepackage{slashed}

\usepackage{mathtools}

\usepackage{tablefootnote}

\usepackage{colortbl}
\definecolor{LightCyan}{rgb}{0.88,1,1}

\usepackage[dvipsnames]{xcolor}
\usepackage{hyperref}
\hypersetup{colorlinks=true, citecolor=blue}

\usepackage{authblk}

\usepackage[style=phys,articletitle=false,biblabel=brackets,chaptertitle=false,pageranges=false,url=false,doi=false,eprint=true,isbn=false,backend=bibtex]{biblatex}

\DeclareFieldFormat{eprint:arxiv}{%
  	\ifhyperref{%
  		\iffieldundef{eprintclass}{%
  			\mkbibbrackets{\href{http://arxiv.org/abs/#1}{\nolinkurl{arXiv:#1}}}
  	}{%
  		\mkbibbrackets{%
    		\href{http://arxiv.org/abs/\thefield{eprintclass}/#1}{\nolinkurl{arXiv:\thefield{eprintclass}/#1}}}
	}
    }{}
}

\DeclareFieldFormat{doi/url-link}{%
      {#1}%
}

\DeclareBibliographyDriver{article}{%
  \usebibmacro{bibindex}%
  \usebibmacro{begentry}%
  \usebibmacro{author/translator+others}%
  \setunit{\labelnamepunct}\newblock
  \usebibmacro{title}%
  \newunit
  \printlist{language}%
  \newunit\newblock
  \usebibmacro{byauthor}%
  \newunit\newblock
  \usebibmacro{bytranslator+others}%
  \newunit\newblock
  \printfield{version}%
  \newunit\newblock
  \printtext[doi/url-link]{%
    \usebibmacro{journal+issuetitle}%
    \newunit
    \usebibmacro{byeditor+others}%
    \newunit
    \usebibmacro{note+pages}%
    \newunit\newblock
    \iftoggle{bbx:isbn}
      {\printfield{issn}}
      {}%
    \setunit{\addspace}%
    \printfield{year}%
  }%
  \setunit{\addspace}%
  \iffieldundef{pages}
    {%
    }%
    {}%
  \usebibmacro{doi+eprint+url}%
  \newunit\newblock
  \usebibmacro{addendum+pubstate}%
  \setunit{\bibpagerefpunct}\newblock
  \usebibmacro{pageref}%
  \newunit\newblock
  \usebibmacro{related}%
  \usebibmacro{finentry}%
}

\usepackage[top=2.5cm, bottom=2.5cm, left=2.5cm, right=2.5cm]{geometry}

\begin{document}
\title{\textbf{Solving DWF Dirac Equation Using Multisplitting Preconditioned Conjugate Gradient}}
\author[1]{Duo Guo}
\author[1]{Robert Mawhinney}
\author[1]{Jiqun Tu}

\affil[1]{Department of Physics, Columbia University, New York, NY 10027}

\date{}
\maketitle

\begin{abstract}
	We show that using the multisplitting algorithm as a preconditioner for conjugate gradient inversion of the domain wall Dirac operator could effectively reduce the inter-node communication cost, at the expense of performing more on-node floating point operations. This method could be useful for supercomputers with far more on-node flops than inter-node communication bandwidth.
\end{abstract}

\section{Introduction}
\subsection{Situation and Motivation}

The cost of lattice QCD simulations with dynamical fermions are dominated by the inversion of Dirac operator, in both the ensemble generation phase and measurement phase. Conjugate gradient(CG) proves to be a stable algorithm to solve these linear equations but the convergence rate is limited by the condition number of the Dirac operator, which becomes large as we start to work with lattices with physical pion mass.  

For the measurement phase various eigen-space methods, including \texttt{EigCG}\cite{Stathopoulos2010} and \texttt{Lanczos}\cite{YSaad1980} method, have been developed successfully to speed up the inversion. Low lying eigenvectors of the Dirac operator are generated and the condition number is effectively reduced to improve the convergence rate of CG. The cost of eigenvector generation is well compensated for the fact that for measurement phase we usually solve Dirac equations with the same Dirac operator with multiple right hand side(RHS). The cost of each of these solves are greatly reduced, therefore it gives a reduction in total computation cost.

This is, however, not the case for the ensemble generation phase. The Dirac operator keeps evolving as the gauge field keeps evolving, and for each of these Dirac operators only a few(usually 1 or 2) Dirac equations are to be solved, thus render it not worthwhile to generate the low lying eigenvectors.

On the other hand, the development of supercomputers has greatly increase the floating point operations per second(flops) on each processor or node. Modern lattice simulations usually divide the gauge field and pseudo-fermion field into sub-fields that are stored and computed locally on different nodes of a large parallel computer. This further improves the total floating point operation capability but requires communications between the nodes. Local computation and communication operates alternately for plain CG. On some latest machines\footnote{For example the SUMMIT machine at ORNL: \url{https://www.olcf.ornl.gov/olcf-resources/compute-systems/summit/}} the communication has become the bottleneck: for CG the local computation speed is almost $10\sim 20$ times faster than the communication speed and the large local flops could not be utilized.

Subsequently for the ensemble generation phase we need a different algorithm that consumes more local computation but less communication than plain CG. Specifically we will focus on (generalized) domain wall fermion(DWF), whose formulation makes it different from, say, Wilson fermion. We will discuss this in the method section.

\subsection{Domain Decomposition Algorithm}

In \cite{Luscher2004} a \textit{domain decomposition} algorithm is proposed for Wilson fermion. The processor grid(node grid) is checkerboarded into black and white nodes. Local inversions are first performed on all black nodes with all pseudo-fermion fields on white nodes, which essentially server as the Dirichlet boundary for the black nodes inversions, are fixed. After the fields on black nodes are updated through the inversion the same is performed on all white nodes, with fields on black nodes serve as the Dirichlet boundary. Iterations of this procedure itself constitute a converging algorithm. It is found in \cite{Luscher2004} that using this procedure as a preconditioner for CG gives faster convergence rate.

Note is that for Wilson fermion the Dirac operator only involves nearest neighbors and the operator is hermitian. These conditions make it possible to only checkerboard the node grid into two colors of black and white.

\section{Method}
\subsection{Multisplitting Algorithm}

In \cite{OLeary1985} a \textit{multisplitting} algorithm is proposed for solving generic linear system. Compared to domain decomposition algorithm, multisplitting does not require checkerboarding. It uses the \textit{current} solution outside of each node as the Dirichlet boundary to perform local inversions. This acts as one iteration. After each iteration, the new boundaries/solutions are communicated to ready the next iteration.

\begin{figure}[]
	\centering
	\includegraphics[]{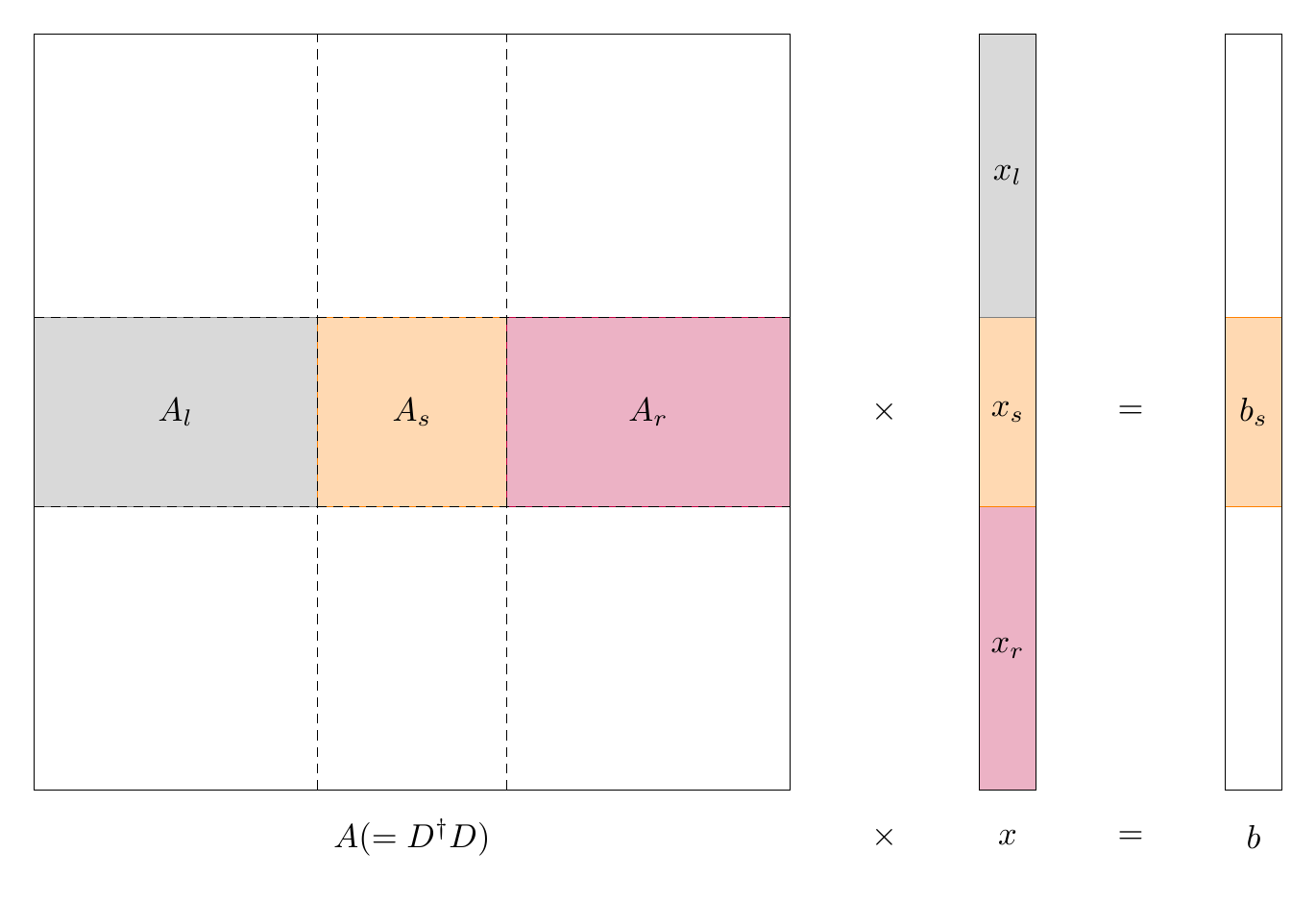}
	\caption{Decomposition of the matrix $A$, the solution vector $x$ and the right-hand-side vector $b$ into local parts on a node.}\label{fig:ms_dec}
\end{figure}
Following \cite{Jezequel2012}, suppose the equation to be solved is $Ax=b$. For a \textit{particular} node the matrix $A$ and vectors $x$ and $b$ are decomposed according to figure \ref{fig:ms_dec}, where $x_s$ and $b_s$ are the part that is locally stored. The original equation turns into
\begin{equation}
	A_sx_s+A_lx_l+A_rx_r=b_s.
\end{equation} 
The $A_lx_l+A_rx_r$ involves off-node vector parts and is calculated before each iteration via communication. Then for each of iterations the algorithm solves the equation 
\begin{equation}\label{eq:ms}
	A_sx_s=b_s-A_lx_l-A_rx_r
\end{equation}
locally for $x_s$ on this local node. The updated solution $x_s$ will then be communicated to other nodes that need it. Apparently this whole procedure could be done concurrently on all nodes once the communication work to calculate $A_lx_l+A_rx_r$ is done.

The domain decomposition algorithm and this algorithm treat the Dirichlet boundary differently and therefore the orders of local inversions are different. The former requires checkerboarding and the later does not.

\subsection{Domain Wall Fermion}

\begin{figure}[]
	\centering
	\includegraphics[width=\textwidth]{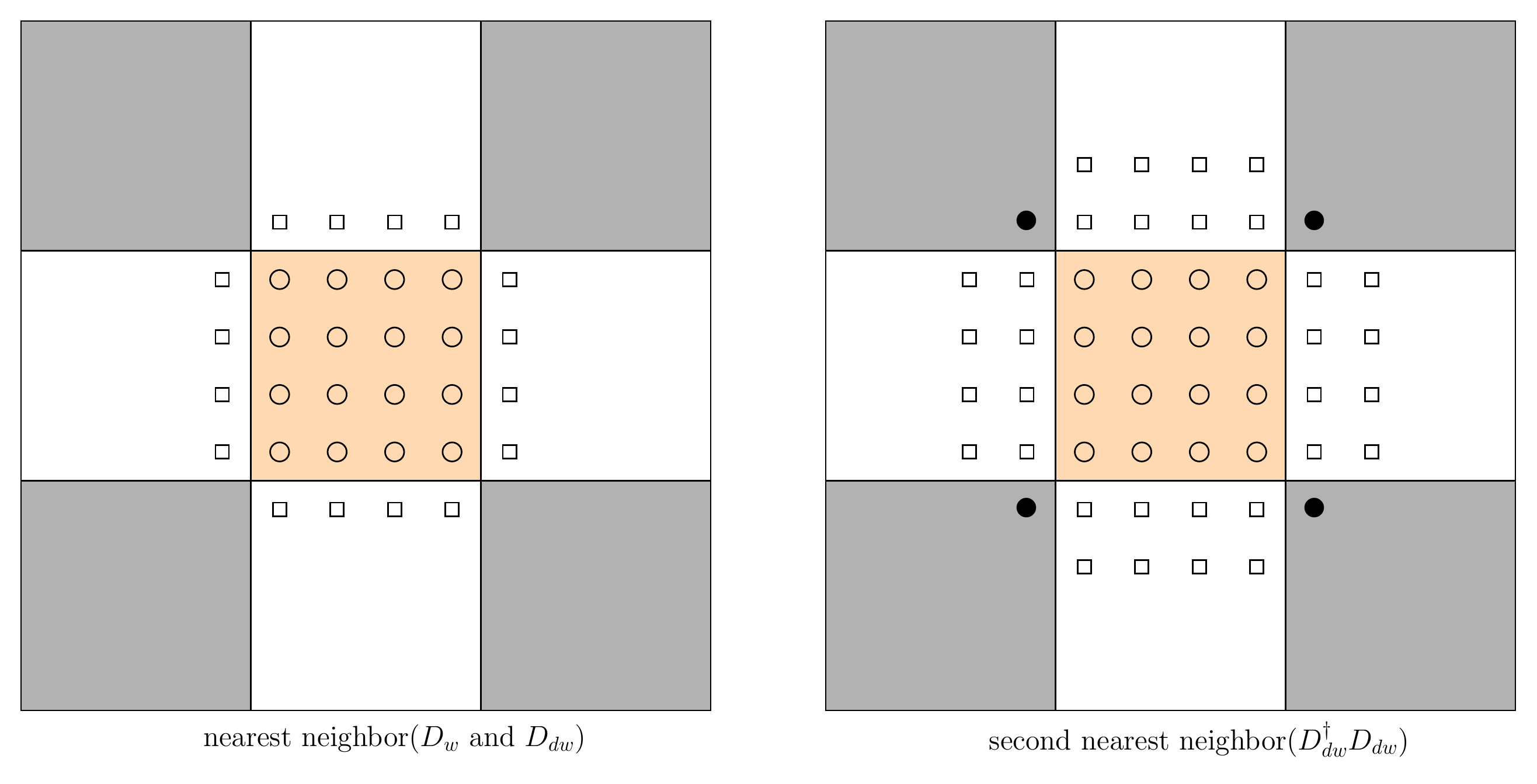}
	\caption{The $D^\dag_{dw}D_{dw}$ operator involves second nearest neighbor.}\label{fig:neighbor}
\end{figure}

Domain wall fermion(DWF)\cite{Jansen1996} is essentially a tensor product of Wilson fermion with large mass and a fictitious fifth dimension. The Dirac operator $D_{dw}$ for DWF is not hermitian so we could not solve $D_{dw}x=b$ directly. Instead we solve
\begin{equation}
	D^\dag_{dw}D_{dw}x=D^\dag_{dw}b.
\end{equation}
The operator $D^\dag_{dw}D_{dw}$ involves not only the nearest neighbors but also the second nearest neighbors. As shown in figure \ref{fig:neighbor}, the $\bullet$ points, which are not present in the $D_{w}$/$D_{dw}$ case, forbid us from using the domain decomposition algorithm as it is proposed in \cite{Luscher2004}, where checkerboarding with only two colors(black and white) are used. A checkerboarding of $2^4=16$ colors would probably make the same algorithm work but this larger factor of $16$ would probably compromise any gains we might get as well.

The multisplitting algorithm does not require checkerboarding. We tried implementing the algorithm with DWF directly but the convergence is slow, quite similar to the situation in \cite{Luscher2004}. Instead we use it as a preconditioner in CG.

\subsection{Preconditioned CG}

\begin{figure}[]
	\centering
	\includegraphics[width=\textwidth]{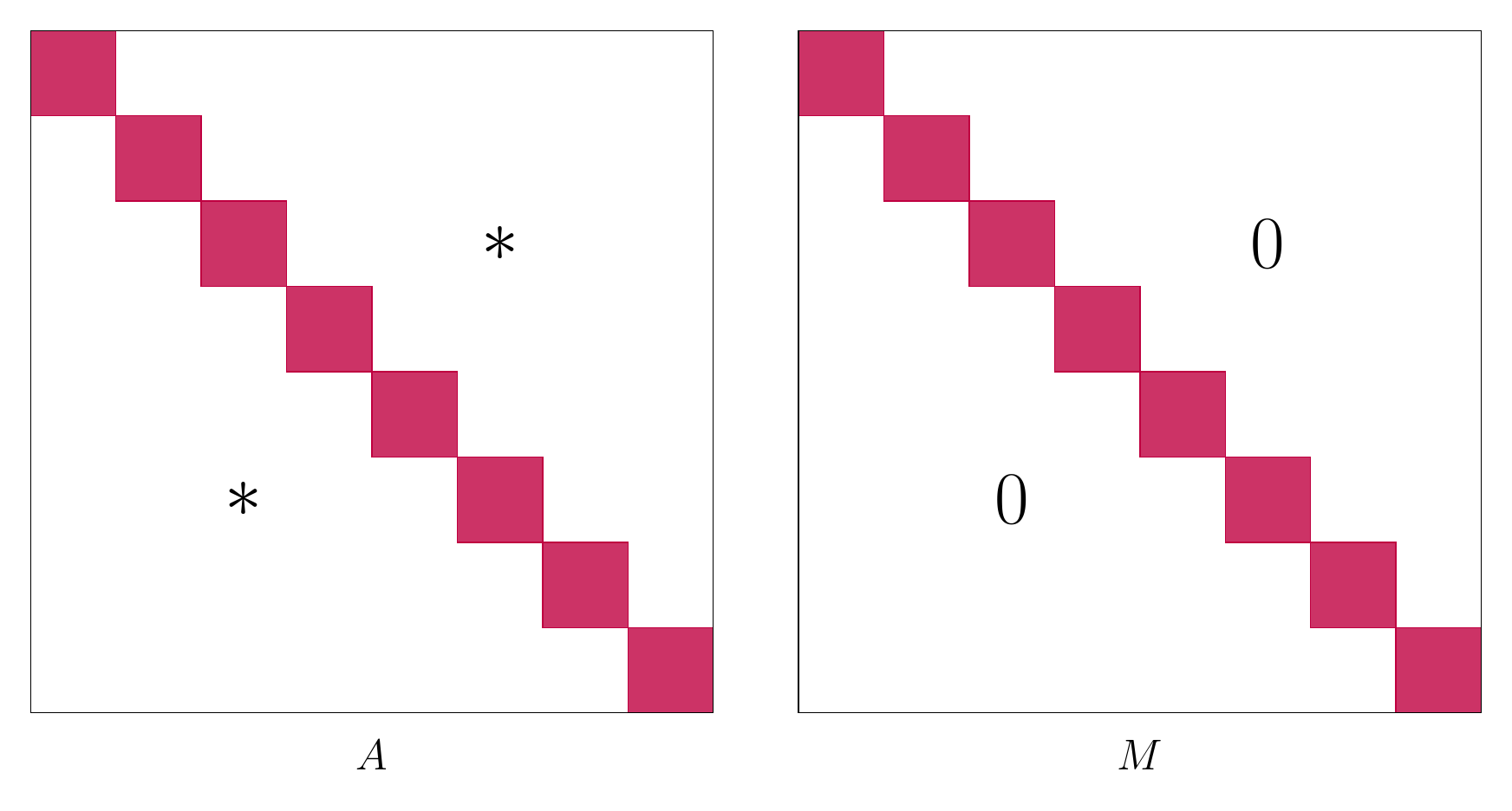}
	\caption{Structure of $A$ and $M$. All off-diagonal elements in $M$ are set to zero.}\label{fig:mprec}
\end{figure}

The operator we are solving is $A=D^\dag_{dw}D_{dw}$. We feed in the right hand side and start with a zero guess solution in (\ref{eq:ms}). This is equivalent to solving
\begin{equation}
	A_sx_s=b_s
\end{equation}
on each node locally. Define 
\begin{equation}
	M=\bigoplus_s A_s,\ s=\mathrm{node\ index}.
\end{equation}
$M$ is the preconditioner for $A$ in our preconditioned CG. As shown in figure \ref{fig:mprec} $M$ has the same local block diagonal entries with $A$ but all off-diagonal elements are set to zero. This structure allows us to solve $M^{-1}b_s$ locally by local (plain) CG.

$M$ as a matrix is fixed, hermitian and positive definite, as long as $A$ has these properties, therefore it is an \textit{exact} preconditioner. In practical numerical simulation, however, it is not possible to achieved exact solution when solving for $M^{-1}$. We can only solve to certain precision. As long as this imperfection does not jeopardize the convergence of the overall CG, we could even reduce the computation cost by solve $M^{-1}$ in a sloppy way. We could use lower floating point precision in the actual numerical implementation and only do a few iterations when solving $M^{-1}$. We will call the iterations spent on locally solving $M^{-1}$ as \textit{local iterations}.

\section{Results}

We implement this multisplitting precondtioner for CG and use the method on the $32^3\times 64$ lattice with physical pion mass on a $128$ KNL machine. As mentioned in the previous section we change the number of local iterations to achieve the maximum numerical efficiency. Compared to CG without preconditioner we achieve a factor of $3$ reduction in global iteration count, which essentially means the same factor of reduction in communication cost, since the $M^{-1}$ solves does not require communication.
\begin{figure}[H]
	\centering
	\includegraphics[]{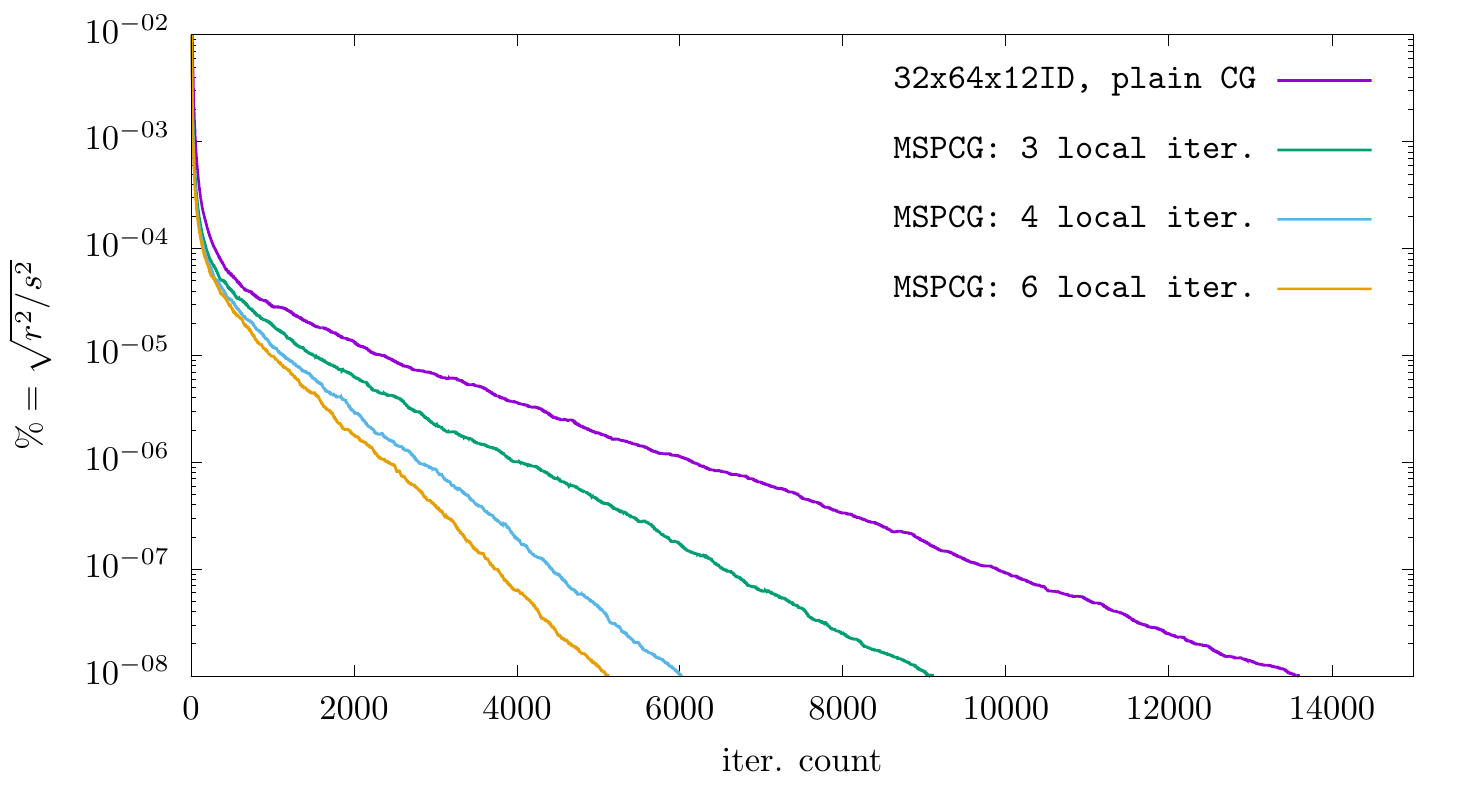}
	\caption{Relative size of global residual as a function of global iteration count. MSPCG refers to multisplitting preconditioned CG.}\label{fig:32ID_N128}
\end{figure}

\section{Discussion and Conclusion}

\subsection{Cost Arithmetic}

Suppose on a parallel machine for plain CG the computation cost is $P$ and the communication cost is $C=\alpha P$. Assume the number of local iterations is $l$ and the reduction factor of global iteration is $g$, then the cost of plain CG is
\begin{equation}
	t_{\mathrm{CG}} = P+\alpha P=(1+\alpha)P
\end{equation}
while the cost of multisplitting preconditioned CG is
\begin{equation}
	t_{\mathrm{MSPCG}} = \left((1+l)P+\alpha P\right)/g.
\end{equation}
Assuming $\alpha=20$, $l=6$ and $g=3$ we would get a factor of $2$ speed up.

\subsection{Condition Number}

We expect the preconditioner $M$ represents the high lying modes of $A$ very well since eigenvectors with high eigenvalues are expected to be local. This preconditioner effectively brings the condition number of the subsequent $AM^{-1}$ matrix down, thus gives a reduction in global iteration count.

This also argues for the idea of solving local $M^{-1}$ in a sloppy way: a few local iterations would get the high modes on each node right, which probably gets the global high modes right as well. We don’t want to do more local iterations to get the low modes on each modes, since those low modes are different from global low modes anyway.

\section{Acknowledgments}

The author wishes to thank Chulwoo Jung, Christopher Kelly and Norman Christ for the suggestions and comments. The numerical implementation of algorithm is done under \texttt{CPS}, \href{https://github.com/paboyle/Grid}{\texttt{Grid}} and \href{https://github.com/waterret/Qlattice}{\texttt{Qlattice}}.

\urlstyle{tt}
\printbibliography

\end{document}